\documentstyle[multicol,prl,aps]{revtex}
\input epsf

\def\etal{{\it et~al.\ }}
\begin{document}
\titlepage
\title{Doubly Enhanced Skyrmions in $\nu =2$ Bilayer Quantum Hall States}
\author{N. Kumada$^{\rm 1}$, A. Sawada$^{\rm 1}$, Z. F. Ezawa$^{\rm 1}$, S. Nagahama$^{\rm 1}$, H. Azuhata$^{\rm 1}$\\
K. Muraki$^{\rm 2}$, T. Saku$^{\rm 2}$, and Y. Hirayama$^{\rm 2}$}

\address{$^1$Department of Physics, Tohoku University, Sendai 980-8578, Japan}

\address{$^2$NTT Basic Research Laboratories, 3-1 Morinosato-Wakamiya, Atsugi, Kanagawa 243-0198, Japan}
\date{Version: \today}
\vspace{5mm}

\maketitle
\begin{abstract}
By tilting the samples in the magnetic field, we measured and compared the Skyrmion excitations in the bilayer quantum Hall (QH) state at the  Landau-level filling factor $\nu =2$ and in the monolayer QH state at $\nu =1$.
The observed number of flipped spins is $N_s=14$ in the bilayer system with a large tunneling gap, and $N_s=7$ in the bilayer system with a small tunneling gap, while it is $N_s=7$ in the monolayer system.
The difference is interpreted due to the interlayer exchange interaction.
Moreover, we have observed seemingly preferred numbers $N_s=14,7,1$ for the flipped spins by tilting bilayer samples.

\end{abstract}

\vspace*{-7mm}
\begin{multicols}{2}\narrowtext
\setcounter{page}{1}

The bilayer quantum Hall\,(QH) state attracts much recent attention especially at the Landau-level filling factor $\nu =2$.
At this filling the competition between the tunneling and the Zeeman effect leads to interesting physics.
Indeed, a phase transition has been observed between the spin polarized state and the spin unpolarized state, as was  revealed by magnetotransport measurements \cite{SawadaPRL,SawadaPRB}, light scattering spectroscopy \cite{PellegriniA,PellegriniB} and capacitance spectroscopy \cite{Khrapai}.
The existence of the interlayer coherence has been pointed out \cite{SawadaPRB,EzawaPRL} in the $\nu =2$ spin unpolarized bilayer QH state.
Moreover, some theoretical works suggest a new phase, that is a canted antiferromagnetic state \cite{DasSarma}.
In the $\nu =2$ spin polarized bilayer QH state, electrons in each layer tend to configure the monolayer $\nu =1$ QH state separately \cite{SawadaPRL}, which is referred to as the compound state.
The state is realized at the balanced point when the total electron density $n_t$ is high enough \cite{SawadaPRL,EzawaE}.
It is important to explore whether there is any difference between the excitations in the compound $\nu =2$ state and the simple monolayer $\nu =1$ QH state, since it yields a deep insight into the role of the interlayer Coulomb and tunneling interactions.

In the monolayer $\nu =1$ QH state, the Coulomb interaction makes the excitation energy much larger than the expected single particle Zeeman energy.
Provided the Zeeman effect is small, the lowest energy charged excitations are spin textures known as Skyrmions \cite{Sondhi,Barrett,Schmeller,Aifer,Bayot,Maude,Leadleyc,Melinte}: They are characterized by the number of flipped spins $N_s$.
It is determined from the measurements of the activation energy by tilting the sample in the magnetic field with keeping the perpendicular component $B_{\perp }$ fixed.
The in-plane magnetic field $B_{\parallel }$ couples to the system only through the Zeeman energy.
The dependence of the excitation energy gap $\Delta $ on the total magnetic field $B_{\rm tot}$ is

\begin{equation}
\Delta =\Delta _{0,s}(B_{\perp })+N_s|g^{\ast }|\mu _{B}B_{\rm tot}.
\label{sky}
\end{equation}
The activation energy $\Delta $ is determined from the temperature dependence of the magnetoresistence: $R_{xx}=R_{0}\exp(-\Delta /2T)$.
The first term $\Delta _{0,s}$ is the contribution to the gap from the non-Zeeman effect, $g^{\ast }$ is the effective gyromagnetic ratio and $\mu _B$ is the Bohr magneton.
From this equation, the number of flipped spins $N_s$ is determined by $\partial \Delta/\partial (|g^{\ast }|\mu _{B}B_{\rm tot})$.

In this Letter, we investigate the Skyrmion excitations in the compound $\nu =2$ state.
We compare the activation energy of the bilayer $\nu =2$ QH state in different $\Delta _{\rm SAS}$ samples and of the induced monolayer $\nu =1$ QH state.
Here, the induced monolayer state is constructed by emptying the electrons in one layer in the same double quantum well sample.

Three samples with different barrier height but the same barrier width were grown by molecular beam epitaxy.
They consist of two GaAs quantum wells of width 200\,\AA\ separated by a 31\,\AA\ thick barrier of Al$_{x}$Ga$_{1-x}$As ($x=0.3$, 0.33 and 1).
We label them \#10.9, \#7.6 and \#1 according to their tunnel energy gap $\Delta _{\rm SAS}$; 10.9\,K, 7.6\,K and 1\,K, respectively. 
($\Delta _{\rm SAS}$ of the highest barrier sample can not be resolved in the Shubnikov-de Hass measurement. We estimate $\Delta _{\rm SAS}=1$\,K from a self consistent calculation.)
A unique structure of the samples \#10.9 and \#1 is that the modulation doping is made only on the front layer, and the back layer electron is fully field-induced through an $n^{+}$-GaAs layer acting as a back gate \cite{Muraki}.
Hence, one can control the electron density of the back layer $n_b$ from 0 to 1.2$\times 10^{11}$\,cm$^{-2}$ by adjusting the back gate bias from 0 to 1.2\,V, while the electron density of the front layer $n_f$ is controlled by adjusting a Ti/Au front Schottky gate bias.
This sample structure enables us to realize easily the balanced bilayer system ($n_f=n_b$) and the monolayer system ($n_{f}\neq 0$, $n_b=0$).
On the other hand, the modulation doping is made on both layers in the sample \#7.6.
The low temperature mobility of the samples \#10.9 and \#1 is 2$\times 10^{6}$\,cm$^2/$Vs with the electron density of 2$\times 10^{11}$\,cm$^{-2}$, while that of the sample \#7.6 is 0.3$\times 10^{6}$\,cm$^2/$Vs with the electron density of 2.6$\times 10^{11}$\,cm$^{-2}$.

Measurements were performed with the samples mounted in a mixing chamber of a dilution refrigerator.
The magnetic field with maximum 13.5\,T was applied to the samples.
Standard low-frequency ac lock-in techniques were used with currents 20\,nA to avoid heating effects.
The samples mounted on a goniometer with the superconducting stepper motor \cite{GONO} rotate into any direction in the magnetic field in unit of $0.05^{\circ }$.

\begin{figure}[!bth]
\epsfxsize=80mm
\epsfbox{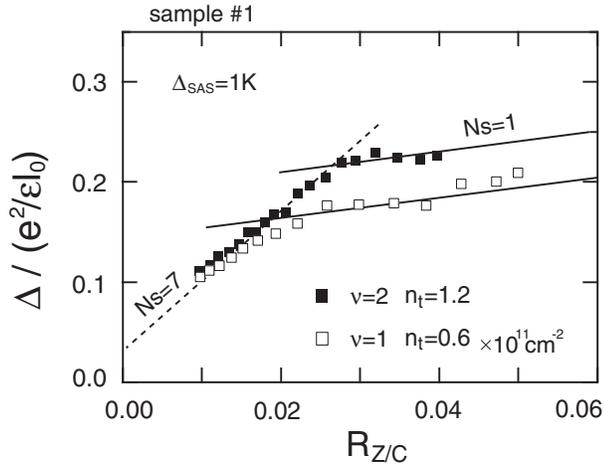}
\caption{\small
The activation energy by tilting the sample \#1 of the $\nu =1$ and $\nu =2$ QH states as a function of the normalized Zeeman energy $R_{Z/C}$.
The solid squares are for the bilayer $\nu =2$ QH state at the balanced point ($n_f=n_b$).
The open squares are for the induced monolayer $\nu =1$ QH state.
For comparison, we have drawn the lines with $N_s=7$\,(dashed) and $N_s=1$\,(full).}
\label{sample1}
\end{figure}

Figure\,\ref{sample1} presents the results of measurements by tilting the sample \#1 ($\Delta _{\rm SAS}=1$\,K) in the magnetic field.
The activation energy divided by the Coulomb energy is plotted vs.\,the Zeeman energy divided by the Coulomb energy $R_{Z/C}=|g^{\ast }|\mu _BB/(e^2/\epsilon \ell _0)$, where $\ell _0=\sqrt{\hbar /eB_{\perp}}$ is the magnetic length.
We used the effective gyromagnetic ratio $g^{\ast }=-0.46$ and the dielectric constant $\epsilon =12.9$. 
Each datum set starts from the magnetic field normal to the two dimensional plane ($B_{\rm tot}=B_{\perp }$).
The solid squares are for the compound $\nu =2$ state, and the open squares are for the induced monolayer $\nu =1$ QH state.

As Fig.\,\ref{sample1} shows, in the $\nu =2$ data at $n_t=1.2\times 10^{11}$\,cm$^{-2}$, the activation energy initially rises quickly as the total magnetic field increases, where the number of flipped spins $N_s=7$ is found.
At $R_{Z/C}=0.027$, the slope changes suddenly and we obtain $N_s=1$ for $R_{Z/C}\geq 0.027$.
The induced monolayer $\nu =1$ data at $n_t=0.6\times 10^{11}$\,cm$^{-2}$ share all these properties except for the inflection point.

\begin{figure}[!bth]
\epsfxsize=80mm
\epsfbox{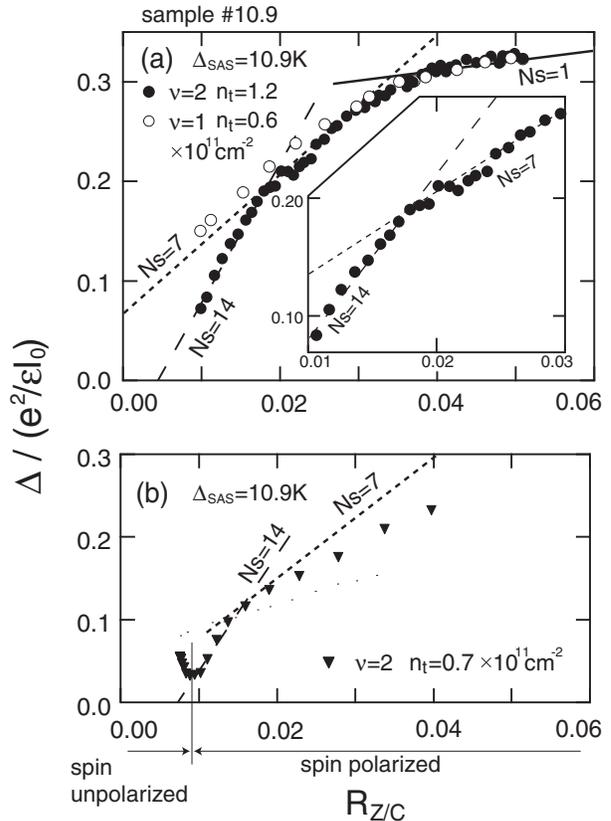}
\caption{\small
The activation energy by tilting the sample \#10.9 of the $\nu =1$ and $\nu =2$ QH states as a function of the normalized Zeeman energy $R_{Z/C}$.
The solid marks are for the bilayer $\nu =2$ QH state at the balanced point.
The open marks are for the induced monolayer $\nu =1$ QH state.
(The inset shows the data of the bilayer $\nu =2$ QH state around the inflection point from $N_s=14$ to $N_s=7$.)
For comparison, we have drawn the lines $N_s=14$\,(long-dashed), 7\,(dashed) and 1\,(full).
In (b), a spin polarized (compound) state is realized for $R_{Z/C}\geq 0.009$.}
\label{sample10}
\end{figure}

Figure\,\ref{sample10} presents the results of measurements by tilting the sample \#10.9 ($\Delta _{\rm SAS}=10.9$\,K) in the magnetic field.
The solid circles are for the compound $\nu =2$ state, and the open circles are for the induced monolayer $\nu =1$ QH state.

From the $\nu =2$ data at $n_t=1.2\times 10^{11}$\,cm$^{-2}$ in Fig.\,\ref{sample10}(a), the number of flipped spins $N_s=14$ is derived for $R_{Z/C}\leq 0.018$.
It is to be emphasized that $N_s=14$ is precisely twice as many as $N_s=7$, which is the one observed in the sample \#1 at the same $R_{Z/C}$ value.
$N_s$ changes from 14 to 7 at $R_{Z/C}=0.018$, and finally changes to 1 at $R_{Z/C}=0.033$.
On the contrary, the induced monolayer $\nu =1$ data at $n_t=0.6\times 10^{11}$\,cm$^{-2}$ show a  similar behavior to the induced monolayer $\nu =1$ data in the sample \#1 in Fig.\,\ref{sample1}.

We have so far focused on the spin polarized (compound) state at $\nu =2$, which is realized at higher density.
We now study the state at $\nu =2$ at low density, which is a spin unpolarized state \cite{EzawaE}.
The data at $\nu =2$ with $n_t=0.7\times 10^{11}$\,cm$^{-2}$ in Fig.\,\ref{sample10}(b) show a rapid decrease up to $R_{Z/C}=0.009$, as in the bilayer $\nu =1$ QH state \cite{Marphy}, which is a signal of interlayer coherence spontaneously developed on the $\nu =2$ QH state \cite{SawadaPRB}.
The new feature for the $\nu =2$ QH state is that they start to increase at $R_{Z/C}=0.009$.
The behavior of the activation energy for $R_{Z/C}\geq 0.009$ is the one peculiar to the compound $\nu =2$ state \cite{SawadaE}.
Namely, the state is stable only at the balanced point and the activation energy increases as the sample is tilted.
It is interpreted \cite{SawadaE,Min} that a phase transition occurs from the spin unpolarized (coherent) state to the spin polarized (compound) state, because $\Delta _{\rm SAS}$ decreases as $B_{\parallel }$ increases \cite{Hu}.
Also in this compound state, we have derived $N_s=14$ at $0.009\leq R_{Z/C}\leq 0.018$, but the number of flipped spins at higher $R_{Z/C}$ is slightly smaller than 7.

\begin{figure}[!bth]
\epsfxsize=80mm
\epsfbox{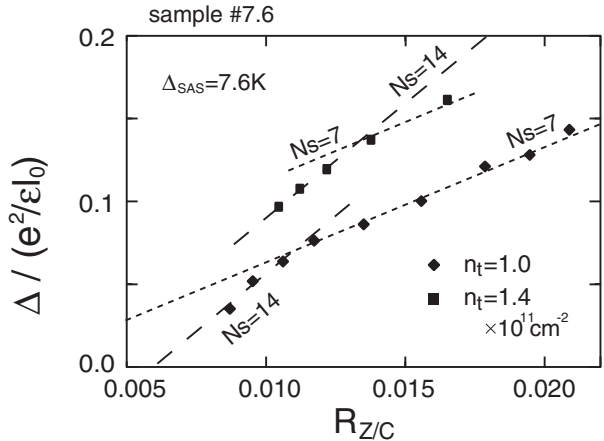}
\caption{\small
The activation energy with tilting the sample \#7.6 of the $\nu =2$ QH states at the balanced point as a function of the normalized Zeeman energy $R_{Z/C}$.
For comparison, we have drawn the lines $N_s=14$\,(long-dashed) and 7\,(dashed).}
\label{sample7}
\end{figure}

In Fig.\,\ref{sample7}, we show the data in the sample \#7.6 ($\Delta _{\rm SAS}=7.6$\,K), whose mobility $0.3\times 10^{6}$\,cm$^2$/Vs is one order lower than that of the sample \#10.9.
These data show a similar behavior to that in the sample \#10.9, and we also recognize $N_s=14$ and 7 from the slope in this sample.

It is essential to compare the compound $\nu =2$ state at $n_t=1.2\times 10^{11}$\,cm$^{-2}$ to the monolayer $\nu =1$ QH state at $n_t=0.6\times 10^{11}$\,cm$^{-2}$.
We have tuned the density so that this compound $\nu =2$ state is made of two monolayer $\nu =1$ states at $n_t=0.6\times 10^{11}$\,cm$^{-2}$.
On one hand, in the sample \#1 (Fig.\ref{sample1}) with $\Delta _{\rm SAS}=1$\,K, the excitation with 7 spin flip was observed both in the compound $\nu =2$ state and in the induced monolayer $\nu =1$ QH state.
On the other hand, in the sample \#10.9 (Fig.\ref{sample10}(a)) with $\Delta _{\rm SAS}=10.9$\,K, the excitation with 14 spin flip was observed in the compound $\nu =2$ state at $R_{Z/C}\leq 0.018$, but 7 spin flip in the induced monolayer $\nu =1$ QH state.
From these results, it is natural to conclude that the excitation with 14 spin flip (7 spin flip) occurs when the tunneling interaction is large (small).

\begin{figure}[!bth]
\begin{center}
\epsfxsize=79mm
\epsfbox{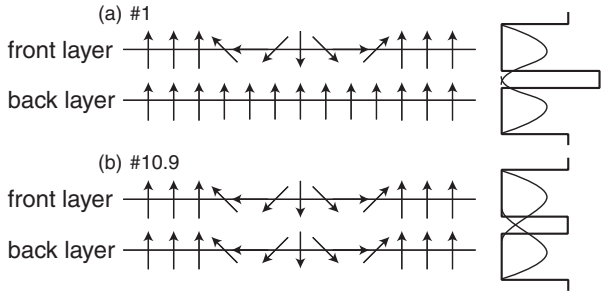}
\end{center}
\caption{\small
Schematic diagram of spin flip in the compound $\nu =2$ state.
An arrow represents the direction of a spin.
(a) In the sample \#1 with a small tunneling gap, spin excitations are identical to those in the monolayer $\nu =1$ QH state.
(b) In the samples \#10.9 with a large tunneling gap, spin excitations in one of the layers affect those in the other layer due to the interlayer exchange interaction.
A Skyrmion-Skyrmion pair will be excited if the interaction is ferromagnetic.
We have illustrated the overlap of the wave functions in the right side.}
\label{intersky}
\end{figure}

Let us elucidate the difference of spin excitations in these two samples with small and large tunneling gaps.
Note that the difference does not originate in the direct interlayer Coulomb interaction since it is identical between the two samples.
The monolayer $\nu =1$ QH state is a QH ferromagnet, where all spins are aligned in a single direction not only by the Zeeman effect but also by the intralayer Coulomb exchange interaction.
We now consider the compound $\nu =2$ state, which is made of two monolayer QH ferromagnets.
The two layers are independent when the tunneling gap is essentially zero.
Hence, we obtain spin excitations identical to those in the monolayer QH state in the sample \#1.
However, a large tunneling gap implies a large overlap of the wave functions, which makes the interlayer exchange interaction operate.
Consequently, a spin flip in one of the layers affects the spin texture in the other layer.
It is natural to expect that Skyrmions are doubly created on the two layers in the sample \#10.9 due to the interlayer exchange interaction.
It is an intriguing problem whether the interlayer exchange interaction induces the ferromagnetic or antiferromagnetic interaction.
The former enhances a Skyrmion-Skyrmion pair (Fig.\ref{intersky}).
The mechanism is quite akin to that in the $\nu =2$ interlayer-coherent phase, where the experimental data \cite{SawadaPRL} is interpreted by a pair excitation of Skyrmions \cite{EzawaPRL}.
On the other hand, the antiferromagnetic interaction \cite{antiDas} will enhance a Skyrmion-anti-Skyrmion pair.
However, the present magnetotransport experiment is unable to tell the type of the interaction.

We next question why the compound $\nu =2$ state in the sample \#10.9 shows a similar behavior to that in the monolayer QH state for larger $R_{Z/C}\geq 0.018$.
This will presumably be because $\Delta _{\rm SAS}$ is decreased by the in-plane magnetic field $B_{\parallel }$ and the interlayer exchange interaction is suppressed.

The energy and the size of the Skyrmion are theoretically estimated at $\nu =1$ as \cite{Sondhi,EzawaJ}
\begin{equation}
\frac{\Delta }{e^2/\epsilon \ell _0}\simeq \sqrt{\frac{\pi }{32}}+\frac{3\beta }{4\kappa }-\Gamma _{\rm{offset}},
\label{skyd}
\end{equation}
where $\beta $ represents the strength of the Coulomb energy which depends on the sample parameter such as the layer thickness ($\beta =3\pi ^2/64$ for a large Skyrmion in an ideal planer system), and $\kappa $ is the size of the Skyrmion,
\begin{equation}
\kappa \simeq \frac{\beta ^{1/3}}{2}\left\{R_{Z/C}\ln \left(\frac{\sqrt{2\pi }}{32R_{Z/C}}+1\right)\right\}^{-1/3}.
\label{skys}
\end{equation}
The offset $\Gamma _{\rm{offset}}$ may be due to the impurities in the sample.
The number of flipped spins $N_s=\partial (\Delta/(e^2/\epsilon \ell _0))/\partial R_{Z/C}$ depends smoothly on $R_{Z/C}$.
The monolayer data due to Schmeller \etal \cite{Schmeller} are fitted reasonably well by this formula \cite{EzawaPRL}.

On the contrary, our experimental data can not be explained by a smooth change of the activation energy.
Clearly there exist preferred numbers of flipped spins $N_s=14,7,1$.
The same behaviors are seen in samples with very different mobility ($2\times 10^{6}$\,cm$^2$/Vs and $0.3\times 10^{6}$\,cm$^2$/Vs).
Because of this fact, the preferred number is not due to the impurity effect.
It must have a different origin, a reminiscence of the magic number momentum in a quantum dot system where electrons configure polygonal pattern originated from the Coulomb interaction \cite{Kuramoto,Maksym}.
It is plausible that the non-Zeeman term $\Delta _{0,s}(B_{\perp })$ of the activation energy makes a local minimum at these preferred numbers to make virtual Wigner Crystal locally.
A further experiment is needed to confirm the conjecture.
It is intriguing that sudden change of $N_s$ were observed also in other experiments \cite{Maude,Leadleyc,Melinte}.

In conclusion, we have measured the activation energy of the $\nu =2$ QH state at the balanced point ($n_f=n_b$) and the $\nu =1$ QH state at the monolayer point ($n_{f}\neq 0$, $n_b=0$) by tilting the samples.
We used three samples with different $\Delta _{\rm SAS}$ and mobility.
In the samples \#10.9 and \#7.6, the excitation with 14 spin flip was observed in the compound $\nu =2$ state, which is twice of 7 spin flip observed in the induced monolayer $\nu =1$ QH state.
In the sample \#1, on the contrary, the number of flipped spins is the same in the compound $\nu =2$ state and in the induced monolayer $\nu =1$ QH state.
We have argued that this difference is due to the interlayer exchange interaction.
As another prominent feature, we have posited seemingly preferred numbers $N_s=14,7,1$ for flipped spins.

We thank H. Aoki for useful discussions.
The research was supported in part by Grant-in-Aids for the Scientific 
Research from the Ministry of Education, Science, Sports and Culture 
(10203201,10640244, 11125203, 11304019), Mitsubishi Foundation, CREST-JST and NEDO ``NTDP-98" projects.


\end{multicols}
\end{document}